\documentclass{ws-p10x7}

\def\be{\begin{equation}}
\def\ee{\end{equation}}

\begin{document}

\title{Reconstruction of Supersymmetric Theories at High Energy Scales}

\author{W.~Porod}

\address{Inst.~de F\'\i sica Corpuscular (IFIC), CSIC, E-46071--Val\`encia,
      Spain \\E-mail: porod@ific.uv.es}

\twocolumn[\maketitle\abstract{
We have studied the reconstruction of supersymmetric
theories at high scales by evolving the fundamental parameters
from the electroweak scale upwards. Universal minimal supergravity and
gauge mediated supersymmetry breaking have been taken as representative
alternatives. Pseudo-fixed point structures require the low--energy
boundary values to be measured with high precision.}]


{\bf 1.} When supersymmetry is discovered and its spectrum of particles
and their properties are measured, the mechanism of electroweak symmetry
breaking must be determined. This will be related to the reconstruction
of the supersymmetric theory at high energy scales. This problem will be
addressed in this report. In particular, we will explore the bottom-up 
approach, which is the most unbiased method in this context.
First indications might already be given
by the particle spectrum and various experimental signatures.
We assume that precision measurements at a high luminosity $e^+ e^-$ linear 
collider\cite{tesla}  (LC) are available. 
More details and an extended list of references are given
in Ref.\cite{Blair00}.

As representative examples we study minimal supergravity 
(mSUGRA)\cite{nilles} 
and gauge mediated supersymmetry breaking (GMSB)\cite{giudice}. 
mSUGRA is characterized by a GUT scale $M_U$ of $O(10^{16}$~GeV) 
where the gauge 
couplings unify. $M_U$ is also 
the scale
of supersymmetry breaking which is parameterized by a common gaugino mass 
parameter $M_{1/2}$, a common
scalar mass parameter $M_0$ and a common trilinear coupling $A_0$ between
sfermions and Higgs bosons.
GMSB is characterized  by a messenger
scale $M_m$ in the
range between $\sim 10$ TeV and $\sim 10^6$~TeV. 
In this scenario the mass parameters
of particles carrying the same gauge quantum numbers squared are universal.
The regularity for scalar masses would be observed at the scale $M_m$
while the gaugino mass parameters should unify at 1-loop order 
at the GUT scale $M_U$ as in the mSUGRA case.
%


%
\begin{figure*}
\setlength{\unitlength}{1mm}
\begin{picture}(60,68)
\put(-0.5,-83){\mbox{\epsfig{figure=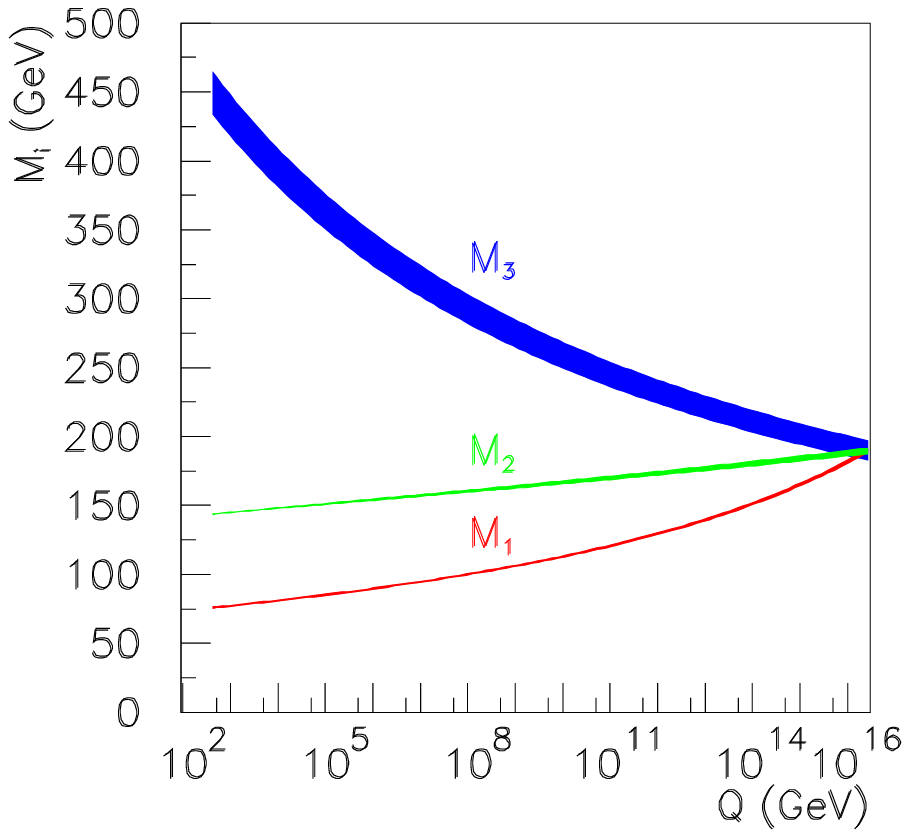,height=16.5cm,width=15cm}}}
\put(71,-83){\mbox{\epsfig{figure=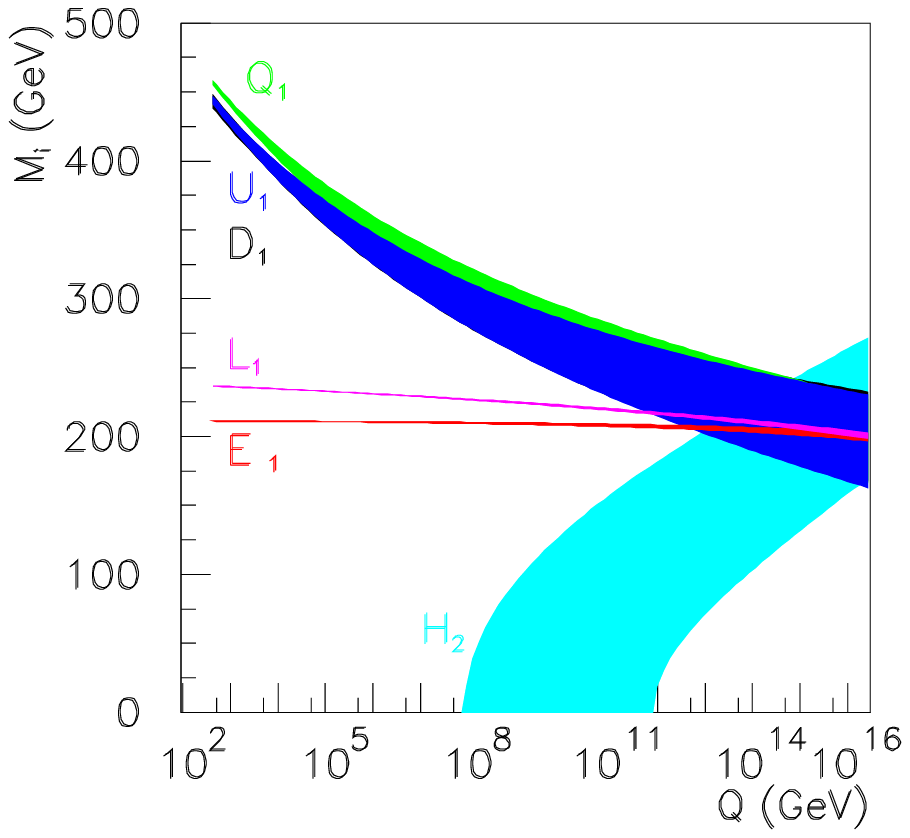,height=16.5cm,width=15cm}}}
\put(-1.5,69){\mbox{\bf (a)}}
\put(70,69){\mbox{\bf (b)}}
\put(45,60){\mbox{ mSUGRA}}
\end{picture}
\caption{{\bf mSUGRA:} {\it  Evolution of 
(a) gaugino and (b) sfermion mass parameters in the
bottom--up approach. The mSUGRA point probed is characterized by the
parameters $M_0 = 200$~GeV, 
$M_{1/2} = 190$~GeV, $A_0$ = 550~GeV, $\tan \beta = 30$, 
and $\mathrm{sign}(\mu) = (-)$.
[The widths of the bands indicate the 95\% CL.]
}}
\label{fig:sugra}
\end{figure*} 
{\bf 2.} The mSUGRA point we have analyzed in detail, is
characterized by: $M_{1/2} = 190$~GeV, $M_0 =
200$~GeV, $A_0$ = 550~GeV, $\tan \beta = 30$, and $\mathrm{sign}(\mu)
= -$. The modulus of $\mu$ is calculated from the requirement of
radiative electroweak symmetry breaking. We have checked the compatibility
of this point with $b \to s \gamma$ \cite{CLEO}  and the 
$\rho$-parameter\cite{Drees90}. 
We have used two-loop renormalization equations\cite{RGE2} to define the 
parameters at the electroweak scale where we also have calculated the
threshold effects\cite{bagger}. 

The parameters provide the experimental observables, including the
supersymmetric particle spectrum and production cross sections. These
observables are endowed with errors as expected from a future LC
experiment\cite{blair2}.
The analysis of the entire particle spectrum except the gluino
requires LC energies 
up to 1 TeV and an integrated luminosity of about 1 ab$^{-1}$.
The errors given in  Ref.\cite{blair2} are scaled
in proportion to the masses of the spectrum. Moreover,
they are inflated conservatively for particles  that
decay predominantly to $\tau$ channels, according to typical reconstruction
efficiencies such as given in Ref.\cite{Nojiri96}.
Typically the relative errors expected from measurements at a Linear
Collider are $O(10^{-3})$ for weakly interacting 
particles and $O(10^{-2})$ for strongly interacting particles (for more
details see Table 1 in \cite{Blair00}).  
For the cross sections we use purely statistical errors, 
assuming a conservative reconstruction efficiency of 20\%.
In the case of the gluino we assume that LHC can measure its mass within an
error of 10~GeV \cite{atlas}.

These observables together with the corresponding errors are interpreted as 
the  experimental data from the experiment and they are used to reconstruct
the underlying fundamental
parameters. These parameters are evaluated in the bottom-up
approach to the grand unification scale within the given errors. 
The results for the evolution of the mass parameters to the
GUT scale $M_U$  are shown in Fig.~\ref{fig:sugra}.
The left-hand side (a) of the figure presents the evolution of the gaugino
parameters $M_i$ which apparently is under excellent control, 
as is the extrapolation
of the slepton mass parameter in Fig.~\ref{fig:sugra}(b). The accuracy
deteriorates for the squark mass parameters and for the Higgs mass parameter
$M_{H_2}$. This can be understood after inspecting the corresponding RGEs.
In case of $M_{\tilde Q_{1}}$ the parameter receives a rather large 
contribution from $M_3$
when renormalized from the high scale down to the electroweak scale: 
The error on $M_3$ is large compared to $M_{1,2}$ as can be seen from 
Fig.~\ref{fig:sugra}a) and is in turn the source of the error on the
squark mass.  
In case of $M_{H_2}$ large Yukawa couplings lead to a pseudo-fixed point
behaviour implying a rather weak dependence of the electroweak parameter
on the original parameter at the high scale.

\begin{figure}
\setlength{\unitlength}{1mm}
\begin{picture}(60,62)
\put(-2,-83){\mbox{\epsfig{figure=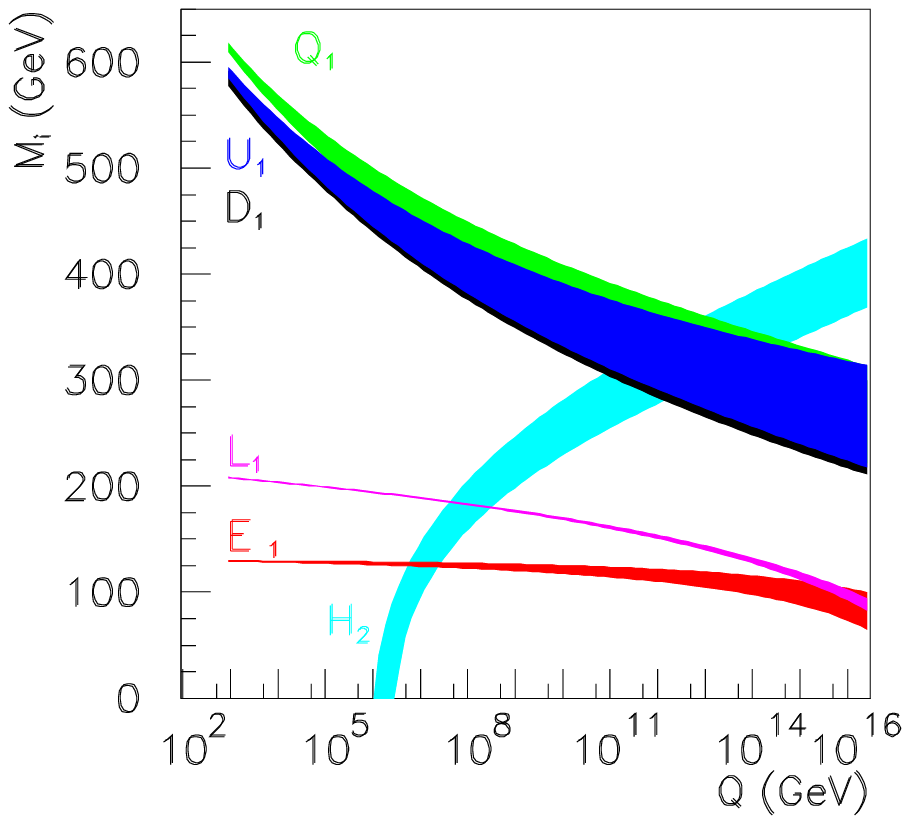,height=16.5cm,width=15cm}}}
\put(35.5,8){\vector(0,1){6}}
\put(45,60){\mbox{GMSB}}
\end{picture}
\caption{{\bf GMSB:} {\it Evolution of sfermion 
mass parameters in the bottom--up approach. The GMSB point has been 
chosen as $M_m = 2 \cdot 10^5$~TeV, 
$\Lambda = 28$~TeV, $N_5 = 3$, $\tan \beta = 30$, 
and $\mathrm{sign}(\mu) = (-)$. 
[The widths of the bands indicate the 95\% CL.]
}} 
\label{fig:gmsb}
\end{figure}

Inspecting Fig.~\ref{fig:sugra}(b) leads to the conclusion that the 
top-down approach eventually may generate an incomplete picture. 
{\it Global} fits based on mSUGRA without allowing for deviations
from universality, are dominated by $M_{1,2}$ and the slepton mass
parameters.  Therefore, the structure of the theory in the squark sector
is not scrutinized stringently at the unification scale
in the top-down approach. 
By contrast, the bottom-up approach demonstrates very clearly the extent
to which the theory can be tested at the high scale.

{\bf 3.} The analysis has been repeated  for gauge 
mediated supersymmetry breaking.
Regularity among particles carrying the
same gauge quantum numbers squared, should  be observed in 
the evolution of mass parameters  
at the messenger scale $M_m$. 
The evolution of the sfermion mass parameters of the 
first/second generation and the Higgs mass parameter $M_{H_2}$
is presented in Fig.~\ref{fig:gmsb}. It is obvious that $M_{H_2}$
approaches the 
mass parameter for the
left-chiral sleptons $M_{L_1}$ 
at the GMSB scale which is indicated by the arrow. 
Moreover, the
figure demonstrates clearly that GMSB will not be confused  
with  the  mSUGRA scenario  as no  more  regularity 
can be observed at the GUT scale $M_U$.

{\bf 4.} The model--independent reconstruction of the fundamental
supersymmetric theory at the high scale, the grand unification scale $M_U$ in
supergravity or the intermediate scale $M_m$ in gauge mediated supersymmetry 
breaking,  appears  feasible. Regular patterns can be observed by evolving the 
gaugino and scalar mass parameters from the measured values at the electroweak
scale to the high scales. The necessary high accuracy  requires in 
addition to the LHC input values high--precision LC values.
The future experimental input from LC is  particularly important if the
universality at the GUT scale is (slightly) broken. 

\section*{Acknowledgments}

I am grateful to G.A.~Blair and P.M.~Zerwas for a very interesting and 
fruitful collaboration.
This work has been supported 
by the Spanish 'Ministerio de Educacion y Cultura' under 
the contract SB97-BU0475382, by Spanish DGICYT grants PB98-0693, and by the 
EEC under the TMR contract ERBFMRX-CT96-0090.

\end{document}